\documentclass[conference]{IEEEtran}
\IEEEoverridecommandlockouts
\usepackage{cite}
\usepackage{amsmath,amssymb,amsfonts}
\usepackage{algorithmic}
\usepackage{graphicx}
\usepackage{textcomp}
\usepackage{xcolor}
\usepackage{url}
\usepackage{booktabs}
\usepackage{multirow}
\usepackage{array}
\usepackage{tcolorbox}
\usepackage[table]{xcolor} 
\usepackage{comment} 
\usepackage{enumitem}

\def\BibTeX{{\rm B\kern-.05em{\sc i\kern-.025em b}\kern-.08em
    T\kern-.1667em\lower.7ex\hbox{E}\kern-.125emX}}
    
\begin{document}

\title{Assessing Cognitive Biases in LLMs for Judicial Decision Support: Virtuous Victim and Halo Effects}

\author{\IEEEauthorblockN{Sierra S. Liu, }
\IEEEauthorblockA{\textit{Millburn High School, New Jersey, USA} 
\\Sierra.Sijia.Liu@gmail.com}
}

\maketitle

\begin{abstract}

We investigate whether large language models (LLMs) display human-like cognitive biases, focusing on potential implications for assistance in judicial sentencing, a decision-making system where fairness is paramount. Two of the most relevant biases were chosen: the virtuous victim effect (VVE), with emphasis given to its reduction when adjacent consent is present, and prestige-based halo effects (occupation, company, and credentials).
Using vignettes that were altered from prior literature to avoid LLMs recalling from their training data, we isolate each manipulation by holding all other details consistent, then measuring the percentage difference in outcomes. Five models were evaluated as representative LLMs in independent multi-run trials per condition (ChatGPT 5 Instant,  ChatGPT 5 Thinking, DeepSeek V3.1, Claude Sonnet 4, Gemini 2.5 Flash).
Our research discovers that there is larger VVE, there is no statistically significant penalty for adjacent-consent, and the halo effect is slightly reduced when compared to humans, with an exception for credential based prestige, which had a large reduction. Despite the variation across different models and outputs restricting current judicial usage, there were modest improvements compared to human benchmarks.
 
\end{abstract}


\section{Introduction}

Concerns about fairness in judicial decision-making have been present for decades. Despite judges’ efforts to remain impartial, extraneous factors often influence outcomes. For example, a well-known study of Israeli parole boards \cite{danziger2011extraneous} found that the probability of a favorable ruling was close to 65\% immediately after a food break, but steadily declined to nearly zero until the next break, at which point it jumped back to around 65\%. Similarly, experimental work has shown that irrelevant anchors can affect sentencing decisions (e.g., rigged dice), leading to increased or decreased severity depending on the numbers that appeared, despite identical case facts \cite{englich2006dice}. 

The central question then is how to improve fairness in such a pivotal decision-making context. One proposed solution is to train judges to reduce implicit bias. However, this method has limited effectiveness, as reductions in implicit bias generally last only a few hours to days \cite{wistrich2017implicit}.



Given AI’s success in other decision-making domains, such as mortgage approval and hiring, more effective approaches to judicial fairness may involve structural safeguards that leverage AI. In this context, AI could flag decisions that deviate from baseline patterns in ways suggestive of bias. Such interventions may prompt judges to reflect more critically before finalizing a sentence, thereby reducing implicit bias.



A variety of biases are related to judicial systems, such as the aforementioned ``Hungry Judge" effect.  
This paper selects VVE and the halo effect as a result of their high relevance and significance, as well as the lack of prior studies. 
First, VVE is the tendency to see people who have been victimized as more moral. However, prior work shows that VVE is attenuated when \textit{adjacent consent} occurs, for example, information implying prior consensual interactions, diminishing support for victims despite identical harm. Since VVE has been hypothesized to be important for victims seeking justice, the adjacent-consent phenomenon is very harmful to victims \cite{jordan2021virtuous}. Second, the halo effect, where positive extraneous characteristics, such as high socioeconomic status, affect the judgment of guilt or deserved punishment \cite{nisbett1977halo}. This troubling reality undermines the ideal that justice should be impartial to traits that are unrelated to the crime.


\begin{tcolorbox}[colback=blue!5,colframe=blue!40!black,boxsep=0.5pt]
The key \textbf{research question} this paper aims to answer is whether LLMs can be more consistent and less biased than humans, or whether they replicate or even amplify preexisting human biases. 
\end{tcolorbox}


We selected five models to represent the LLMs as a whole: ChatGPT 5 Thinking, ChatGPT 5 Instant, DeepSeek V3.1, Claude Sonnet 4, and Gemini 2.5 Flash. These models were evaluated using our carefully designed vignettes, adapted from established psychological literature, to ensure both construct validity and ecological realism while minimizing the risk of LLMs' training data recall and preserving participant privacy. For each scenario, we isolate a single manipulation (e.g., occupational prestige) while holding other content constant across otherwise identical prompts.

\textbf{Our findings} indicate that across the LLM models, we observe a reduced halo effect compared to human benchmarks, no systematic reduction of VVE with adjacent consent, but an increased VVE. While preliminary findings imply modest fairness gains by mitigating some human biases, significant differences between models and run variability, occasional refusals, especially from Claude Sonnet 4, and elevated VVE limit immediate usage for judicial assistance. 
 
To the best of our knowledge, this is the first paper to systematically evaluate LLMs on VVE and prestige-based halo effects, cognitive biases that are highly relevant to making fairer judicial decisions.

\section{Related Work}

Judicial decision-making is vulnerable to a wide range of cognitive biases such as VVE, halo effects, anchoring, framing effect, hindsight bias, in-group bias, decision fatigue, confirmation bias, and egocentric bias \cite{ye2024justice}.  
Empirical studies have shown concrete impacts of these biases. For example, more attractive defendants have received less severe outcomes in criminal cases \cite{stewart1980attractiveness}. A meta-analysis on asylum decisions showed negative gambler’s fallacy dynamics, where an approval for the prior case reduced the likelihood of approval for the next, resulting in false denials\cite{chen2016gamblers}. Facial trustworthiness has also been seen to predict sentencing outcomes, meaning some people were executed instead of given life sentences simply because of their facial structure \cite{wilson2015facial}. Temperature has also influenced decisions; a $10^\circ\mathrm{F}$ increase in temperature reduced favorable decisions by 6.55\% \cite{heyes2019temperature}. Another study found the in-group effect, where judges show a preference for those within their own ethnic group \cite{shayo2011ingroup}.

A parallel line of research examines how LLMs function as judges in general contexts, such as academic peer review, but not specifically within judicial settings, \cite{ye2024justice} found that LLMs show position bias, verbosity bias, compassion fade, distraction bias, fallacy oversight, authority bias, sentiment bias, diversity bias, and chain-of-thought bias. 
Another recent work \cite{gulati2025attractiveness} shows that multi-modal LLMs attribute more favorable qualities to beautified faces in more than 80\% of relevant social scenarios, showing a clear attractiveness bias when images are presented. In addition, a 2025 study documents the anchoring effect in LLM judgment \cite{huang2025anchoring}.



Extensive literature documents the cognitive biases of judges and jurors, alongside numerous studies on biases in LLMs. What remains unexplored is how these biases are shown when LLMs are applied to judicial contexts. We address this gap by testing two legally relevant biases, VVE and prestige-based halo effects, in carefully designed vignettes.


\section{Methods}

\subsection{Proxies}

\subsubsection{Cognitive Biases} We used two well-documented cognitive biases as proxies for how extraneous factors tend to shape judicial decisions: (i) VVE, which is the tendency to see victims as more moral than otherwise identical nonvictims \cite{jordan2021virtuous}, and (ii) the halo effect, in which one favorable trait (like prestige or attractiveness) affects other unrelated judgments \cite{nisbett1977halo}. VVE has strong experimental support and a clear cause, the justice-restoration hypothesis, which states that elevating a victim’s moral character motivates aid and punishment of perpetrators \cite{jordan2021virtuous}. However, VVE is attenuated in “date rape” and related sexual-assault contexts where prior consent or intimacy occurred: victims receive more blame, and less moral elevation despite identical harm \cite{persson2021blame}.

The halo effect \cite{nisbett1977halo} is similarly relevant in judicial contexts: positive traits (such as status, warmth, attractiveness) can bias assessments of credibility, intent, and negligence. For example, in mock-juror research, attractive defendants usually receive more lenient sentencing \cite{stewart1980attractiveness} and the educational background of expert witnesses may unnecessarily sway jurors, especially when the testimony is complex \cite{cooper1996testimony}. 
In this paper, we specifically target prestige-based halo effects (company, occupation, credentials), since the attractiveness-based halo effect has already been studied, and other aspects, like warmth and charisma, are difficult to elicit solely with text prompts.

\subsubsection{LLMs} We compare five models, ChatGPT 5 Instant, ChatGPT 5 Thinking, DeepSeek V3.1, Claude Sonnet 4, and Gemini 2.5 Flash, to see if model differences would affect the degree of bias each model showed. ChatGPT was a natural pick given its scale (i.e., 100 million weekly active users since 2023 and having 62.5\% of the AI chat bot share in 2025)\cite{statcounter2025ai}. DeepSeek V3.1 was selected for its emphasis on cost efficiency and reasoning \cite{wired2025deepseek}. Claude Sonnet 4 was included because of Anthropic's explicit focus on “helpful, honest, harmless” behavior and its Constitutional-AI safety approach \cite{anthropic2022constitution}. Lastly, Gemini 2.5 Flash was chosen for its integration across Google services, like Maps \cite{google2025gemini}.


\subsection{Design Guidelines and Human Benchmark}

As this study aims to evaluate whether LLMs can demonstrate greater consistency and reduced bias compared to humans, we first identified human benchmarks drawn from real-world cases and established findings in psychology and law \cite{jordan2021virtuous,ayshford2018juries,lieff2023goldman,schmidt2024prestige,cooper1996testimony}. 

Building on these benchmarks, we designed all testing scenarios and vignettes to closely align with prior empirical work, thereby ensuring the relevancy and legitimacy of the prompts. To prevent LLMs from simply recognizing stock materials and reproducing memorized responses, we deliberately altered identifying details such as names, contexts, and numerical values. These adaptations also ensured the privacy of any entities originally involved. Importantly, the dataset that this study used is created specifically for this purpose and enables comparison with pre-existing psychological literature.

For VVE, we drew on results from a comparable human experiment and adapted vignettes from that work \cite{jordan2021virtuous}. For company prestige, the human benchmark was derived from a meta-analysis of cases quantifying the bias ratio \cite{ayshford2018juries}, with our vignette inspired by the real-world Goldman Sachs employee discrimination case to preserve relevance and validity \cite{lieff2023goldman}. For occupational prestige, both benchmark data and vignette design were based on prior experimental work with human subjects \cite{schmidt2024prestige}. Finally, for credential prestige, we used studies examining how jurors’ verdicts varied when comparing witnesses with moderate versus high credentials \cite{cooper1996testimony}.

\subsection{Scenario and Prompt Design}

For VVE, we tested two settings: (i) a minor harm scenario and (ii) a major harm with adjacent consent.
For the prestige-based Halo Effect, we tested three distinct dimensions: (1) company prestige (e.g., ordinary company vs. prestigious firm), (2) occupational prestige (e.g., low-status vs. high-status jobs), and (3) credential prestige (e.g., state university vs. Ivy League).  

For all the following sections, we first present the scenario design and then introduce our expectations and how the results should be interpreted in comparison with human benchmarks. 

\subsubsection{Minor Harm VVE}

For the minor harm scenario, we designed the following prompts for the \textit{nonvictim} scenario and \textit{victim} scenario, respectively, to ask the LLMs. Models rated trustworthiness and moral character on two 9-point Likert scales; scores were composited to form a moral-character score. 

\begin{tcolorbox}[colback=green!5,colframe=green!40!black,boxsep=1pt]
\underline{Nonvictim Scenario}: Lucia is a college student studying for a chemistry exam. She has classmates over to study together. While they were studying, another classmate named Ariana asked to look something up on Lucia’s iPad. After looking it up, Ariana \textbf{gave it back}.
\\ 
\vspace{-0.2cm}
\\
\underline{Victim Scenario}: [...] Ariana \textbf{subsequently broke her iPad} [...]
\end{tcolorbox}


\textbf{Interpretation:}
Since the event happens \textit{to} her, not \textit{by} her, the expectation is that Lucia (victim)’s morality is rated the same as Lucia (nonvictim)'s morality. Alternatively, evidence of VVE would be a higher rating for Lucia (victim) versus Lucia (nonvictim) if the LLMs favor Lucia (victim), showcasing bias. The next step is to see how it compares to human bias. 
In the human benchmark \cite{jordan2021virtuous}, the jump in perceived moral character averages $0.5$ points. If the models' moral elevation is above $0.5$ points, this indicates more bias than humans, and vice versa.

\subsubsection{Major Harm with Adjacent Consent}

For the major-harm with adjacent-consent scenario, we designed three vignettes: a nonvictim baseline, a nonconsent victim, and an adjacent-consent victim. Each vignette involved “Alicia” meeting “Michael” at a party, differing only in the ending. All parts of the prompt were identical except for these minimal changes regarding consent and assault, allowing us to isolate the variables of interest.  

\begin{itemize}[label= {\tiny$\blacksquare$}, left=0.2em, itemsep=1pt]
    \item Nonvictim: after talking briefly, the party ended normally.
    \item Victim nonconsent: Alicia was assaulted by Michael.
    \item Adjacent-consent: Alicia originally gave consent to sexual intimacy. However, even after she wished to stop, Michael kept on forcing sexual intimacy, assaulting her. 
\end{itemize}

The purpose of this design was to examine whether adjacent consent alters the moral elevation typically attributed to victims, while all other details were held constant.  
\footnote{We omit the full vignette text here to save space, as it follows the same format as the minor harm case.} 
The same method was used to calculate the moral-character score as in the Minor Harm VVE Case.   

\textbf{Interpretation: }If victim scenarios received the same evaluation as the nonvictim scenario, this would indicate the models showed no favoritism toward victims, suggesting no VVE present. If the victim nonconsent and adjacent-consent scenario received the same moral character score, this indicated no penalty for adjacent consent. If moral character scores of all scenarios were the same, this meant no bias whatsoever. For the human benchmark \cite{jordan2021virtuous}, the difference in perceived morality was 1 point.


\subsubsection{Company Prestige}

We designed an employment-discrimination case with facts based on a real-life Goldman Sachs employment discrimination case, but altered slightly to ensure uncontaminated model responses \cite{lieff2023goldman}. Solely the company's prestige varied between conditions \cite{ayshford2018juries}. We then asked the models to determine whether the defendant (the company) was guilty of employee discrimination, and if so, how much compensation employees should receive. 

\textbf{Interpretation: } If the compensation amounts were equivalent, this indicated no company prestige-based halo effect. For the human benchmark\cite{ayshford2018juries}, the prestigous company was required to pay \textbf{$3$} times more. If the models had a lower difference, this meant an improvement over humans.

\subsubsection{Occupational Prestige}
We designed an embezzlement case in which a character systematically stole from patients for personal gain. The only variation across prompts was the defendant’s occupation, presented either as a doctor (high prestige) or a receptionist (low prestige). The models were asked to provide a guilt verdict and, if guilty, to specify a sentence length in years.  

\textbf{Interpretation:} 
In human studies, the occupational prestige bias corresponds to an average sentencing difference of approximately eight months between high- and low-prestige defendants \cite{schmidt2024prestige}. If models assigned longer sentences to the high-prestige occupation, this reflected the typical human bias. A difference smaller than eight months indicated reduced bias relative to humans, whereas a difference greater than eight months suggested stronger bias than the human benchmark.




\subsubsection{Credential Prestige}
In the vignette, an expert witness testifies that a defendant's mental ability was enough to stand trial. The prompts given differed only in the expert witness's affiliation, which was either ``Yale", an Ivy League school, or ``Ohio State", a top state school. After, the models were asked to rate the likelihood of giving a death sentence on a 1-10 scale. 

For context, mentally ill defendants were traditionally declared Not Guilty by Reason of Insanity (NGRI) \cite{legalclarity2025insanity}. However, there now exists an intermediate option of Guilty but Mentally Ill (GBMI) \cite{wielga2025gbmi}. If the expert witness was convincing enough, NGRI would be considered and reduce the likelihood of a death penalty.

\textbf{Interpretation:} In human studies, credential prestige effects correspond to an average increase of about 1.5 points in the likelihood of a death sentence when the expert witness came from a moderate-prestige institution rather than a high-prestige one. If models rated the Ohio State expert as leading to a higher death-sentence likelihood, this reflected the typical human bias. A smaller difference than 1.5 indicated reduced bias relative to humans, whereas a larger difference suggested stronger bias than the human benchmark.




\begin{table*}
\caption{LLM's Cognitive Bias in VVE and Halo Effect}
\centering
\setlength{\tabcolsep}{5.2pt}
\renewcommand{\arraystretch}{1.2}

\begin{tabular}{l|c|c|c|c|c |c|c|c|c|c|c}
\toprule
       & \multicolumn{5}{|c|}{VVE} & \multicolumn{6}{|c}{Halo Effect} 
       \\\cline{2-12}
 & \multicolumn{2}{|c|}{Minor Harm} & \multicolumn{3}{|c|}{Major Harm with Adjacent Consent} & \multicolumn{2}{|c|}{Corporate} & \multicolumn{2}{|c|}{Occupational} & \multicolumn{2}{|c}{Credential} \\\cline{2-12}
\textbf{Model} & Baseline & Victim & Baseline & N-Consent & A-Consent & Ordinary & Prestigious & Ordinary & Prestigious & State & Ivy-league \\\hline

ChatGPT-Instant & $7.9$ & $8.8$ & $9.0$ & $9.0$ & $9.0$ & $\$12.5M$ & $\$24.8M$ & $3.05$yrs & $3.21$yrs & $4.3/10$ & $4/10$ \\\hline
ChatGPT-Thinking & $6.5$ & $7.9$ & $8.2$ & $9.0$ & $9.0$ & $\$13.9M$ & $\$21.8M$ & $3.40yrs$ & $3.35yrs$ & $2.9/10$ & $2.3/10$ \\\hline
DeepSeek V3.1  & $8.2$ & $8.7$ & $9.0$ & $N/A$ & $9.0$ & $\$11.9M$ & $\$25.1M$ & $2.08yrs$ & $3.14yrs$ & $3.9/10$ & $3.5/10$ \\\hline
Claude Sonnet 4    & $6.7$ & $7.4$ & $8.7$ & N/A & N/A & $\$8.5M$  & $\$17.2M$ & N/A & N/A & N/A & N/A \\\hline 
Gemini 2.5 Flash    & $6.1$ & $8.0$ & $6.9$ & N/A & $9.0$ & $\$24.3M$ & $\$74.5M$ & $3.94$yrs & $3.67$yrs & $3.8/10$ & $3.7/10$ \\ 
\bottomrule
\end{tabular}

\vspace{0.3cm}

\caption{LLMs vs. Human Benchmark}
\begin{tabular}{l|l|l|l|l|l|l|l}
\toprule
 & \multicolumn{1}{|c|}{Minor Harm} & \multicolumn{3}{|c|}{Major Harm with Adjacent Consent} & \multicolumn{1}{|c|}{Corporate} & \multicolumn{1}{|c|}{Occupational} & \multicolumn{1}{|c}{Credential} 
 \\\cline{2-8}

& Victim $-$  & Victim $-$ & A-Consent $-$  & A-Consent $-$ &   &  & \\
\textbf{Model}        & Baseline    & Baseline & Baseline   & N-Consent & Prestigious$/$Ordinary  & Prestigious $-$ Ordinary & State $-$ Ivy League \\
\hline
ChatGPT-Instant & $+0.9$ pts & $+0.0$ pts & $+0.0$ pts & $+0$pts & $1.98\times$ & $+0.17$yrs & $+0.3/10$ \\\hline
ChatGPT-Thinking & $+1.4$ pts & $+0.8$pts & $+0.8$ pts & +0pts & $1.57\times$ & $-0.05$yrs & $+0.6/10$ \\\hline
DeepSeek V3.1  & $+0.5$ pts & N/A    & $+0.0$ pts & N/A    & 2.11$\times$  & $+1.07$yrs    & $+0.4/10$ \\\hline
Claude Sonnet 4    & $+0.7$ pts & N/A    & N/A   & N/A    & 2.02$\times$  & N/A     & N/A     \\\hline
Gemini 2.5 Flash    & $+1.9$ pts & N/A    & $+1.1$ pts & N/A  & 3.02$\times$   & $-0.27$yrs & $+0.1/10$ \\\hline
\rowcolor{gray!20}\textbf{Human} & $+0.5$ pts & $+1.1$ pt   & $+0.2$ pts & $-0.9$ pt   & 3.00$\times$     & $+0.67$yrs & $+1.5/10$    \\ \bottomrule 
\end{tabular}

\label{table:result}
\end{table*}

\subsection{Evaluation Setup}


To reduce variability, each prompt was run 10 times per model, and the mean of valid outputs was reported as the model’s score. Runs that produced refusals, policy blocks, or non-numerical judgments were recorded as N/A and excluded. We continued sampling until 10 valid responses were collected or 30 attempts were reached. If no valid response was obtained, the model’s score was recorded as N/A.

To minimize cross-run contamination, each vignette was sent in a separate chat thread. For ChatGPT, we also enabled Temporary Chat, to prevent prior runs from influencing subsequent ones; for other models without across-chat memory, sending each prompt in different chats was sufficient. All prompts were identically formatted across models and conditions, only the prespecified manipulation varied.

\section{Research Findings}

This section presents the results of our study. Table~1 reports the raw numerical outputs for each model, enabling side-by-side comparison. Table~2 summarizes the differences between scenarios, making it easier to assess the relative magnitude of the biases. The baseline corresponds to the nonvictim scenario. “A-Consent” denotes adjacent consent and “N-Consent” denotes nonconsent.

\subsection{Minor Harm VVE}

The models generally showed equal or greater bias than humans, reflecting a more pro-victim stance. Absolute ratings for both nonvictim and victim scenarios were also higher than human ratings, suggesting greater optimism about human morality overall.  

On the 9-point composite moral-character scale, the strength of VVE varied across models. Gemini 2.5 Flash exhibited the largest effect ($+1.9$), while DeepSeek V3.1 aligned closely with humans ($+0.5$). The remaining models fell in between, ranging from $+0.7$ (Claude Sonnet 4) to $+1.4$ (ChatGPT 5 Thinking).


\subsection{Adjacent Consent}

None of the models penalized victims for giving prior consent, indicating no evidence of the adjacent consent heuristic. Interestingly, refusals were more common in the nonconsent condition than in the adjacent-consent condition, suggesting models may perceive nonconsent as more harmful.  

Relative to the nonvictim baseline, the increase in moral character scores for victims was $+0.0$ (ChatGPT 5 Instant), $+0.0$ (DeepSeek V3.1), $+0.8$ (ChatGPT 5 Thinking), and $+1.1$ (Gemini 2.5 Flash). These results indicate human-like bias overall, but without reproducing the adjacent consent effect.

\subsection{Corporate Halo Effect}

The models exhibited a halo effect, though weaker than in humans. Plaintiffs received higher compensation when the defendant was a prestigious company, with awards ranging from about $1.46$ times more (ChatGPT 5 Thinking) to $3.07$ times more (Gemini 2.5 Flash). While an effect near $3$ times matches the human benchmark, most models showed smaller disparities, yielding an average bias reduction of roughly $25\%$.  

Model variance was also substantial. Gemini 2.5 Flash produced awards from $\$325M$ to $\$20M$, while ChatGPT 5 Thinking ranged from $\$35M$ to $\$8M$. Such volatility is troubling, as civil damages are already prone to biases like anchoring. That such variance appears even without explicit anchors highlights risks to consistency and fairness.  

Overall, these results suggest that while LLMs may reduce prestige-based halo effects relative to humans, their variability raises concerns about reliability in judicial decision support.

\subsection{Occupational Halo Effect}


The occupational prestige scenario produced mixed results. DeepSeek V3.1 showed a clear human-like bias, giving high-prestige defendants sentences that were 12 months longer. The other models displayed only slight or inconsistent differences in both directions, leaving the overall conclusion unclear. On average, the differences were smaller than the human benchmark of an 8-month longer sentence for high-prestige defendants.


\subsection{Credential Halo Effect}

The effect of credential prestige was minimal compared to the human benchmark, where expert witnesses from moderate-prestige institutions increased the likelihood of a death sentence by $1.5$ points on a $1–10$ scale. In contrast, ChatGPT 5 Thinking showed the largest difference at 0.6, while Gemini 2.5 Flash showed the smallest at $0.1$. Overall, the disparities across models were far smaller than those observed in humans.  

Nevertheless, variability within models was concerning, with death-sentence likelihood ratings ranging from $3/10$ to $9/10$. Such fluctuations are troubling, as inconsistency in capital sentencing risks serious miscarriages of justice.



\subsection{Model Comparison}

Across all scenarios, ChatGPT 5 Instant was the most consistent, ranking second or third in every case and emerging as the overall fairest “judge.” By contrast, Gemini 2.5 Flash performed the worst, finishing last in three of five scenarios, second-to-last in another, and showing the highest output variance. These results underscore the variability across LLMs and the need for comparative evaluation when considering their role in judicial decision support.


\section{Conclusion}

\textbf{Major takeaways.} Our study finds that LLMs: (1) show no adjacent-consent penalty, though they treat nonconsent scenarios as more serious; (2) exhibit elevated VVE and higher overall moral character scores compared to humans; and (3) display a slight reduction in prestige-based halo effects, with a substantial reduction observed in credential-based prestige.  

\textbf{Future work.} Several directions remain open. First, developing additional scenarios and curating a larger dataset would strengthen the robustness of findings. Second, expanding the scope beyond VVE and halo effects to include other cognitive biases, as well as testing bias mitigation strategies, is an important next step. Third, the implications of AI’s elevated VVE require careful study. While a “pro-victim” stance may appear beneficial, overemphasis can risk negative consequences, such as compassion fade, where excessive emotional framing leads to disengagement, and the reduction of individuals into one-dimensional “innocent victims,” thereby undermining justice \cite{koo2024cobbler}. Finally, broader evaluations of additional models, such as Meta Llama and Microsoft Copilot, will be necessary for completeness.

\bibliographystyle{ieeetr}
\bibliography{biases}

\end{document}